\documentclass{pasj00}

\begin{document}
\SetRunningHead{S. NAGATAKI}{Rotating BHs as Central Engines of Long GRBs}

\title{Rotating BHs as Central Engines of Long GRBs:\\ Faster is Better}

\author{Shigehiro \textsc{Nagataki} %
}
\affil{Yukawa Institute for Theoretical Physics, Kyoto University,
 Oiwake-cho Kitashirakawa Sakyo-ku, Kyoto 606-8502, Japan}
\email{nagataki@yukawa.kyoto-u.ac.jp}

%

\KeyWords{gamma rays:bursts --- black hole physics --- relativity ---
supernovae: general --- accretion, accretion disks } 

\maketitle

\begin{abstract}
We performed simulations of collapsars with different 
Kerr parameters $a=$0, 0.5, 0.9, 0.95. It is shown that a more
rapidly rotating black hole is driving a more energetic jet. 
No jet is seen for the case of Schwartzschild black hole case,
while the total energy of the jet is as large as $10^{50}$
erg for a rapidly rotating Kerr black hole case ($a=0.95$).
In order to explain the high luminosity of a GRB, it is concluded 
that a rapidly rotating black hole is favored ('faster is better').
We also find in the case of $a=0.95$ that 
(i) the stagnation region is clearly found in the jet region,  
(ii) the ordered poloidal field lines are seen in the jet,
(iii) the jet region is surrounded by a
'Wall-like' structure that has a higher pressure than the jet region
and contains strong vertical magnetic fields, and 
(iv) the jet is initiated by outgoing Poynting flux from the outer horizon of
the black hole (Blandford-Znajek effect).
The bulk Lorentz factor of the jet is still of the order of unity.
However, energy density of electro-magnetic fields dominates the one 
of rest-mass in the jet. It can be expected
that a relativistic jet will be seen if we perform a
simulation for a longer time scale (of the order of $10-100$ sec).
\end{abstract}

\section{Introduction}

It is still unknown how the central engine of Long Gamma-Ray Bursts
(hereafter it is called as GRBs for simplicity) is working at the
center of massive stars. In other words, we poorly know the engine
that drives the most powerful explosion in the universe. 
Some of the supernovae that associate with GRBs were very energetic.
Their energies were of the order of $10^{52}$ ergs, which is $\sim 10$
times larger than the ones of normal core-collapse supernovae.
This fact strongly suggests that other engine from the one of normal
core-collapse supernovae should be working at the center of the GRB's
progenitors. 

There are some possible scenarios to drive a GRB jet. 
One of the most promising scenarios is the collapsar
scenario (\cite{woosley93}). 
In this study, we investigate the collapsar scenario. 
In the collapsar scenario, a black hole is formed as a result of
gravitational collapse. Also, rotation of the progenitor plays an
important role. Due to the rotation, an accretion disk is formed
around the equatorial plane. After the formation of the accretion disk,
there are several possible stories as below.

One is that a jet-induced explosion along the rotation axis may occur
due to the heating through neutrino anti-neutrino pair annihilation
that are emitted from the accretion
disk~\citep{woosley93,macfadyen99,fryer00,nagataki03a,nagataki07,sekiguchi07,barkov10,harikae10}. 
This is an interesting possibility, although neutrino heating efficiency 
looks small ($0.1-1 \%$). A very careful treatment of neutrino
transfer with conserved-scheme of hydrodynamics will be necessary to
prove that this effect is the key-process of GRB's engine. 

Another is that a jet is driven by extracting rotational energies of the
accretion disk with a help of magnetic fields that pierce 
the disk (Blandford-Payne effect: \cite{blandford82}). This scenario is
also investigated by several
authors~\citep{proga03,proga03b,mizuno04a,mizuno04b,proga05,fujimoto06,nagataki07,suwa07,harikae09}.
This is also a promising possibility. Further study will be necessary
whether a relativistic jet will be launched by this mechanism. 

Recently, the effect of extraction
of rotation energy from the black hole through outgoing Poynting
flux (Blandford-Znajek effect: \cite{blandford77}) is
investigated by using a General Relativistic Magneto-hydro Dynamics
(GRMHD) code~\citep{barkov08,komissarov09a,nagataki09,barkov10}. Energy
extraction from a rotating black hole is a general relativistic
effect, so GRMHD code is necessary to investigate the effect. 
In Barkov and Komissarov (2008), they showed that a jet is launched by
Blandford-Znajek effect using a Kerr black hole with Kerr parameter
($a$=0.9) and a polytrope density profile for a massive star model. 
In their successive papers (Komissarov and Barkov (2009a), Barkov and
Komissarov 2010), they also succeeded to launch a jet by Blandford-Znajek
effect (Kerr Parameter was chosen to be 0.9 in Komissarov and Barkov
2009a, while 0.46 and 0.6 in Barkov and Komissarov 2010) using a
polytrope density profile for a massive star model. 
In Nagataki (2009), it was shown that a jet is launched by
Blandford-Znajek effect (Kerr parameter was chosen to be 0.5) 
using a realistic progenitor model developed
by Woosley and Heger (2006). In Nagataki (2009), it was pointed
out that Blandford-Payne effect may be also working. 

As stated above, in the previous papers, it was shown that a jet is
successfully launched by Blandford-Znajek effect. However, there is no
systematic study how the dynamics depends on the Kerr parameter. It is
true that Komissarov and Barkov (2010) presented results for different
Kerr parameters, but they also changed the amplitude of the initial
magnetic fields. Also, no simulation has been reported for $a=0$, that
is, Schwartzschild black hole case. By performing a simulation for that
case, we can clearly see how effectively the rotating black hole is
working to drive the jet. In this study, we present 4 simulations for
the same initial condition with Nagataki (2009), but with different 
Kerr parameters $a=$0, 0.5, 0.9, 0.95. It is shown clearly that a more
rapidly rotating black hole is driving a more energetic
jet. Especially, in the case of the Schwartzschild case, no jet is
found. That proves that the jet is really driven by the rotating black
hole. 

In section \ref{method}, method of calculation is explained. 
In section \ref{initial}, initial condition is shown. 
Results are given in section \ref{result}. 
Discussion is presented in section \ref{discussion}.
Conclusion is stated in section \ref{conclusion}.

\section{Method of Calculation}\label{method}

In this study,
we use the GRMHD code developed in Nagataki (2009).
Thus we briefly explain the method of calculation here, and please see
Nagataki (2009) for details.  

In Nagataki (2009), we have developed a two-dimensional
GRMHD code following Gammie et al. (2003) and Noble et al. (2006).
We have adopted a conservative, shock-capturing scheme with Harten,
Lax, and van Leer (HLL) flux term (\cite{harten83}) and
flux-interpolated constrained transport technique (\cite{toth00}).
We use a third-order Total Variation Diminishing (TVD)
Runge-Kutta method for evolution in time,
while monotonized central slope-limited linear interpolation method is
used for second-order accuracy in space (\cite{van77}).
2D scheme (2-dimensional Newton-Raphson method) is usually adopted for 
transforming conserved variables to primitive 
variables (\cite{noble06}).

When we perform simulations of GRMHD,
Modified Kerr-Schild coordinate is basically adopted with mass of the black
hole ($M$) fixed where the Kerr-Schild radius $r$ is replaced by the
logarismic radial coordinate $x_1= \ln r$. 
When we show the result, the coordinates are transfered from
Modified Kerr-Schild coordinates to Kerr-Schild ones.
In the following, we use $G=M=c=1$ unit.
$G$ is the
gravitational constant, $c$ is the speed of light, and $M$ is the
gravitational mass of the black hole at the center.

\section{Initial Condition}\label{initial}

The initial condition is also same with Nagataki (2009), but for 
different Kerr Parameters. Thus we explain it briefly here, and 
please see Nagataki (2009) for details. 

The calculated region corresponds to a quarter of the
meridian plane under the assumption of axisymmetry and equatorial
symmetry. The spherical mesh with 256($r$)$\times$ 128($\theta$) grid
points is used for all the computations.
The radial grid is nonuniform, extending from $r=$0.98$r_+$ to 3$\times
10^{4}$ ($r_+ = 1 + \sqrt{1-a^2}$ is the outer horizon, 
and 3$\times 10^{4}$ corresponds to 8.9$\times 10^{9}$cm 
in cgs units) with uniform grids in the Modified Kerr-Schild coordinate.

We adopt the model 12TJ in Woosley and Heger (2006) (as for the
progenitor evolution model, see e.g., Tutukov and Fedorova 2007).
This model corresponds to a star that has 12$M_{\odot}$ initially with
1$\%$ of solar metallicity, and rotates rapidly and does not lose its
angular momentum so much by adopting small mass loss rate. As a result, 
this star has a relatively large iron core of $1.82M_{\odot}$, and
rotates rapidly at the final stage.
We assume that the central part of the star has
collapsed and formed a black hole of 2$M_{\odot}$. 
Since $M=2M_{\odot}$, $r=1$ corresponds to 2.95 $\times 10^5$cm.
We also assume that the gravitational mass of the black hole is 
unchanged throughout the calculation.
In Nagataki (2009), the Kerr parameter, $a$, was assumed to be 0.5,
but in this study we perform simulations for $a=$ 0, 0.5, 0.9, 0.95,
respectively (we name them as Model A,B,C,D, respectively). 

Since 1-D calculation is done for the model 12TJ, we can use the data
directly for the physical quanta on the equatorial plane. As for the
density, internal energy density, and radial velocity, we assume the
structure of the star is spherically symmetric. We also set
$u^\theta=0$ initially. As for $u^\phi$, we extrapolate its value such as
$u^\phi (r,\theta) = u^\phi (r, \pi/2) \times \sin \theta$.

We assume the vector potential $A_\phi \propto
{\rm max} (\rho/\rho_{\rm max} - 0.2,0) \sin^4 \theta  $ where
$\rho_{\rm max}$ is the
peak density in the progenitor (after extracting the central part of
the progenitor that has collapsed and formed a black hole). The field is
normalized so that the minimum value of $p_{\rm gas}/p_{\rm mag} =
10^2$ where $p_{\rm gas}$ is the thermal pressure and $p_{\rm mag}$
is the magnetic pressure.

We use a simple equation of state $p_{\rm gas} = (\Gamma-1)u$ where
$u$ is the internal energy density. We set $\Gamma$=4/3 so that the
equation of state roughly represents radiation gas. 

As for the boundary condition in the radial direction,
we adopt the outflow boundary condition for the inner and outer
boundaries. 
As for the boundary condition in the zenith angle direction, axis of
symmetry boundary condition is adopted for the rotation axis, while
the reflecting boundary condition is adopted for the equatorial plane.
As for the magnetic fields, the equatorial symmetry boundary condition,
in which the normal component is continuous and the tangential component 
is reflected, is adopted.

\section{Results}\label{result}

\begin{figure}
  \begin{center}
    \FigureFile(160mm,160mm){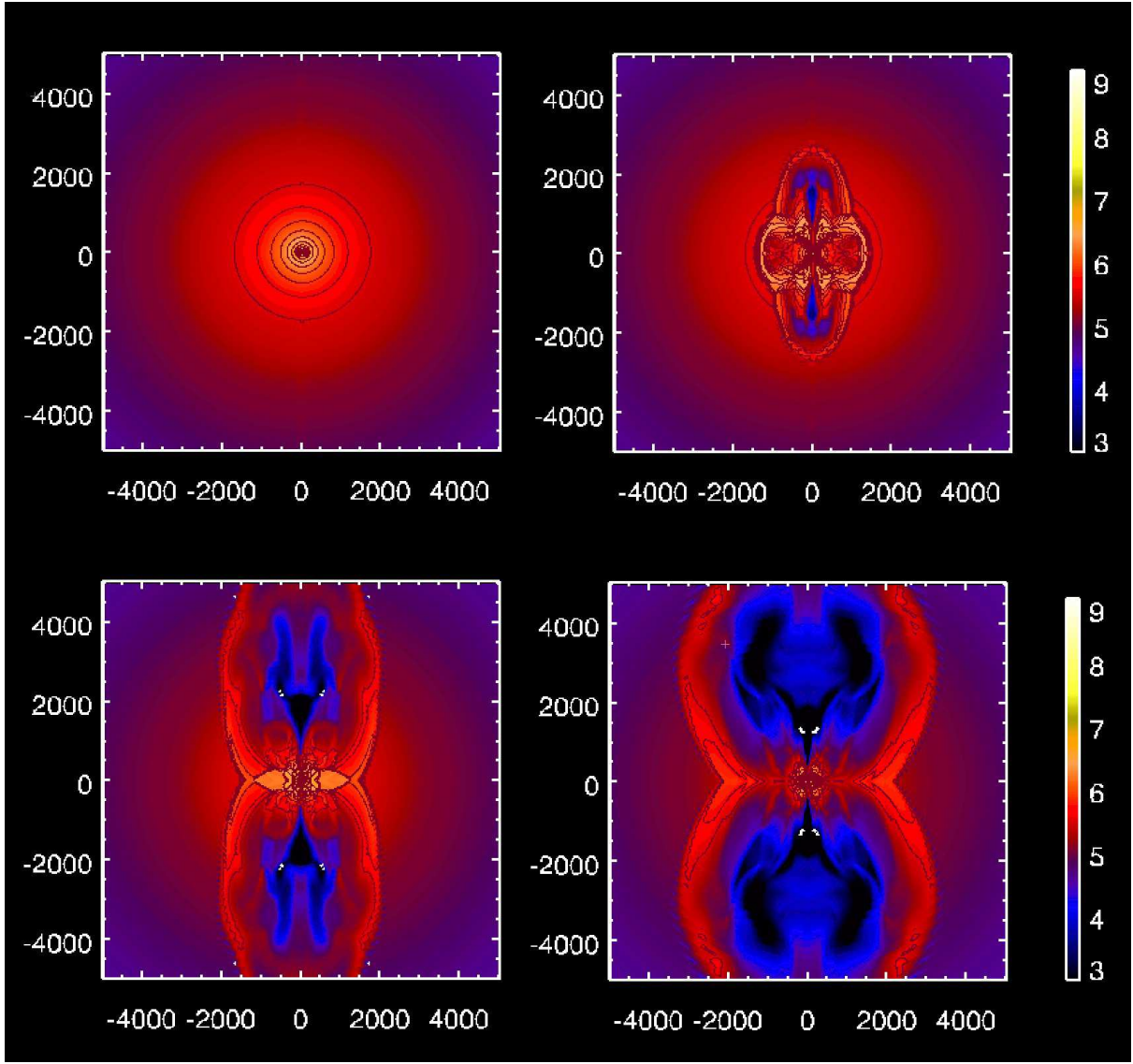}
  \end{center}
  \caption{Contours of rest mass density in logarismic scale for all
models at the same time-slice $t=160000$ (that corresponds to 1.5760 sec).  
Cgs units are used for the rest mass density, while the length
in the vertical/horizontal axes is written in $G=M=c=1$ unit.
$r=1$ and 4000 corresponds to 2.95 $\times 10^5$ cm and 1.18 $\times
    10^9$ cm, respectively.
These results are projected on the (r sin$\theta$, r cos $\theta$)-plane.
Upper left panel shows the state of Model A ($a=0$), upper right panel
    shows the one of Model B ($a=0.5$), lower left panel shows the one
    of Model C ($a=0.9$), and lower right panel shows the one of Model
    D ($a=0.95$).}\label{fig:1}
\end{figure}

In figure \ref{fig:1}, contours of rest mass density
in logarismic scale for all models at the same time-slice $t=160000$ (that
corresponds to 1.5760 sec) are shown.     
Cgs units are used for the rest mass density, while the length
in the vertical/horizontal axes is written in $G=M=c=1$ unit.
$r=1$ and 4000 corresponds to 2.95 $\times 10^5$ cm and 1.18 $\times
10^9$ cm, respectively.
These results are projected on the (r sin$\theta$, r cos $\theta$)-plane.
Upper left panel shows the state of Model A ($a=0$), upper right panel
shows the one of Model B ($a=0.5$), lower left panel shows the one
of Model C ($a=0.9$), and lower right panel shows the one of Model D 
($a=0.95$). It is clearly shown that the rotating black hole drives the
jet (the Schwartzschild black hole cannot drive a jet (Model A),
while a more rapidly rotating black hole is driving a stronger jet). 

\begin{figure}
  \begin{center}
    \FigureFile(80mm,80mm){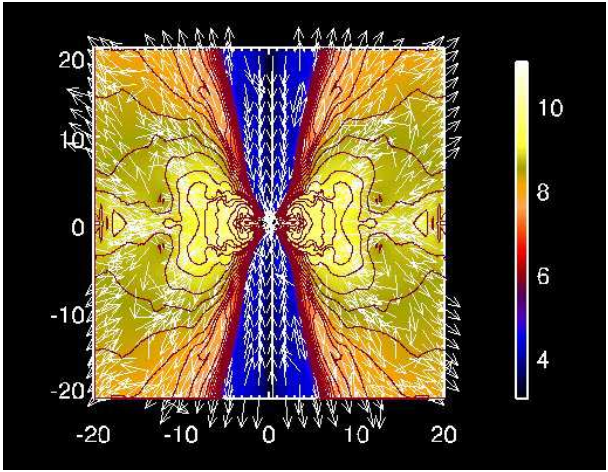}
  \end{center}
  \caption{Same with Figure 1, but for Model D ($a=0.95$) with
    velocity fields. $r=20$ corresponds to 5.9 $\times 10^6$ cm.
Arrows represent the velocity fields ($u^r,u^\theta$).
}\label{fig:2}
\end{figure}
\begin{figure}
  \begin{center}
    \FigureFile(80mm,160mm){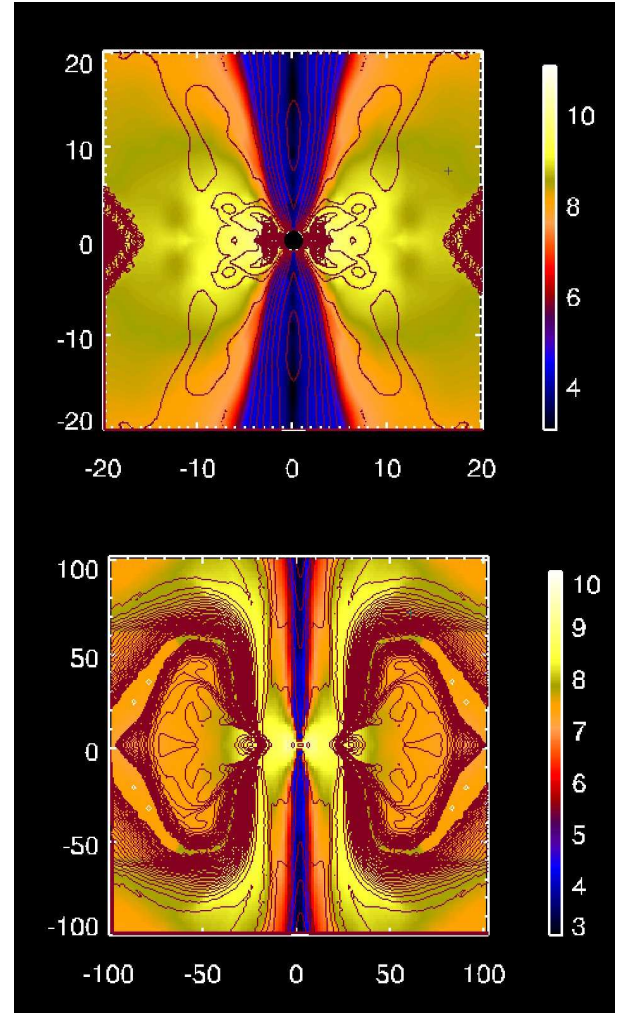}
  \end{center}
  \caption{Same with Figure 1, but for Model D ($a=0.95$) with
    line-contours of the $\phi$ component of the vector potential
($A_\phi$).  
Upper panel shows the central region ($20 \times 20$ in $G=M=c=1$
    unit), while lower panel shows the wider region ($100 \times 100$).
$r=100$ corresponds to 2.95 $\times 10^7$ cm.
}\label{fig:3}
\end{figure}
\begin{figure}
  \begin{center}
    \FigureFile(80mm,80mm){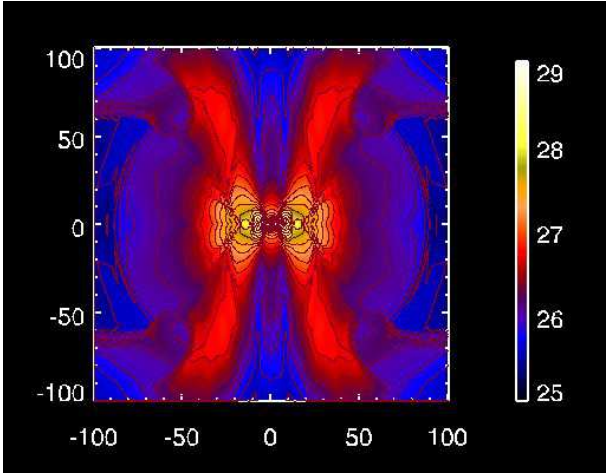}
  \end{center}
  \caption{Contours of total pressure (sum of thermal and magnetic pressure) in
    logarismic scale at $t=160000$ (that corresponds to 1.5760 sec) for
    Model D ($a=0.95$). Cgs units are used for the pressure contours,
    while the length in the vertical/horizontal axes is written in
    $G=M=c=1$ unit. $r=100$ corresponds to 2.95 $\times 10^7$ cm.
}\label{fig:4}
\end{figure}
\begin{figure}
  \begin{center}
    \FigureFile(80mm,80mm){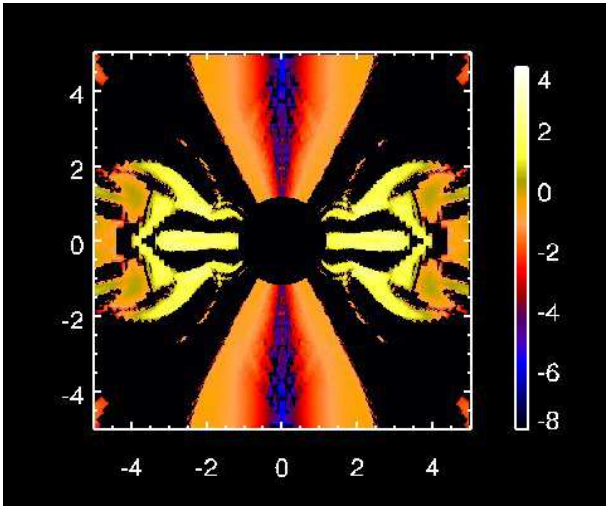}
  \end{center}
  \caption{
Contours of outgoing Poynting flux in
    logarismic scale at $t=160000$ (that corresponds to 1.5760 sec) for
    Model D ($a=0.95$). The unit of the contours is $10^{50}$ erg
    s$^{-1}$ sr$^{-1}$ (see text for the definition in detail),
    while the length in the vertical/horizontal axes is written in
    $G=M=c=1$ unit. $r=1$ corresponds to 2.95 $\times 10^5$ cm.  
    Outer horizon is seen at the center ($r_+=1.312$). 
}\label{fig:5}
\end{figure}
\begin{figure}
  \begin{center}
    \FigureFile(80mm,80mm){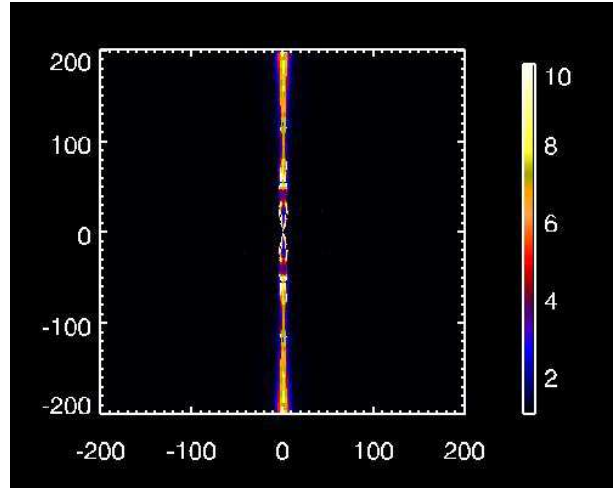}
  \end{center}
  \caption{Contours of the energy flux per
unit rest-mass flux for Model D ($a=0.95$) at $t=160000$.
The contours
represent the bulk Lorentz factor of the invischid
fluid element when all of the internal and magnetic energy are
converted into kinetic energy.  
}\label{fig:6}
\end{figure}

From figure \ref{fig:2} to \ref{fig:6}, we show the results for 
Model D at $t=160000$ (that corresponds to 1.5760 sec).
The length in the vertical/horizontal axes is written in $G=M=c=1$
unit ($r=1,20,100$ correspond to 2.95 $\times 10^5$ cm, 5.9 $\times
10^6$ cm, 2.95 $\times 10^7$ cm, respectively).

In figure \ref{fig:2}, color-contours of rest mass density in logarismic
scale (cgs units) with velocity fields (arrows) are shown. 
It is noted
that stagnation region where the radial velocity becomes positive
(outgoing) from negative (accreting) is seen around $r=15$ in the
jet. In contrast, it is shown later (in figure \ref{fig:5}) that
outgoing Poynting flux is positive from the outer horizon.

In figure \ref{fig:3}, color-contours of rest mass density in logarismic
scale (cgs units) with contours
of the $\phi$ component of the vector potential 
($A_\phi$) are shown. Level surfaces coincide with poloidal magnetic field
lines, and field line density corresponds to poloidal field strength.  
Upper panel shows the central region ($20 \times 20$), 
while lower panel shows the wider region ($100 \times 100$).
In the upper panel, the
ordered poloidal field lines are seen in the jet. In the lower panel,
a 'Wall-like' structure that contains vertical magnetic fields 
is seen.


In figure \ref{fig:4}, contours of total pressure (sum of thermal and
    magnetic pressure) in logarismic scale. Cgs units are used for the
    pressure contours are shown.
It is clearly seen that the 'Wall-like' structure has high pressure:
it is higher than the one in the jet region.

In figure \ref{fig:5}, contours of outgoing Poynting flux ($F_E$ in
    Eq.\ref{BZeq1}) in 
    logarismic scale. 
It is noted that the total energy flux, which is the integrated
    outgoing Poynting flux over the zenith angle, can be written as
\begin{equation}
\dot{E} = 2 \pi \int_0^{\pi} d \theta \sqrt{-g} (-T^r_{\rm EM, \it t}) 
        = 2 \pi \int_0^{\pi}     d \theta F_E,
\label{BZeq1}
\end{equation}
where $g$ and $T^r_{\rm EM, \it t}$ are determinant of the metric and ($r$,$t$)
component of energy-momentum tensor of electro-magnetic fields.
The unit of the contours is $10^{50}$ erg
    s$^{-1}$ sr$^{-1}$.  
    Outer horizon is seen at the center ($r_+=1.312$). 
    It is clearly seen that outgoing Poynting flux is coming out from
    the outer horizon. This figure clearly shows that Blandford-Znajek
    process is working in this system.   
We could see also a powerful outgoing Poynting flux toward the
accretion disk. Partially this may be working for the variability of
the mass accretion rate seen in Nagataki (2009).

In figure \ref{fig:6}, contours of the electro-magnetic field energy flux per
unit rest-mass flux are shown, which represent the bulk Lorentz 
factor of the invischid fluid element when all of the electro-magnetic
field energy are converted into kinetic energy (\cite{nagataki09}). 
Even though the bulk Lorentz factor of the jet is still low (of the
order of unity), the terminal bulk Lorentz factor can be relativistic.

\begin{figure}
  \begin{center}
    \FigureFile(80mm,100mm){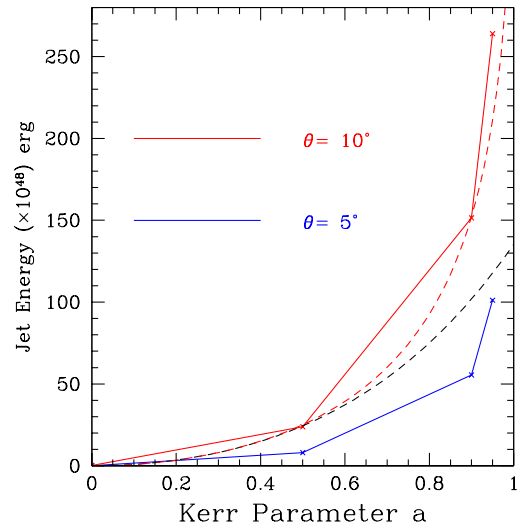}
  \end{center}
  \caption{Plots of the jet energy (see text for the definition in
    detail) for all models at $t=160000$ (that
    corresponds to 1.5760 sec). The unit of vertical axis is $10^{48}$
    erg. Blue curve represents the jet energy within the opening angle
    $\theta = 5^{\circ}$, while red curve represents the one within
    $\theta = 10^{\circ}$. The analytic curves of the BZ-flux formulation of 
    Tchekhovskoy et al. (2010) for the case of monopole solution with
    B=$5 \times 10^{14}$G are shown by the 
    red-dashed curve (forth power 
    of $\Omega_{\rm H}$). The analytic curve of the BZ-flux formulation of 
    Tanabe and Nagataki (2008) for the case of monopole solution with 
    B=$5 \times 10^{14}$G is shown by 
    the black-dashed curve (until forth power of $a$).
}\label{fig:7}
\end{figure}

Finally, in figure \ref{fig:7}, plots of the jet energy for all models
at $t=160000$ are shown.
The definition of the jet energy is:
\begin{equation}
E_{\rm Jet} = 2 \times  2\pi \int_{r_+}^{\infty} dr
\int_{0}^{\theta} d \theta \sqrt{-g} (T^t_{t} - \rho u^0 u_0 ),
\label{GRB3-1}
\end{equation}
where $T^t_{t}$ is the ($t$,$t$) component of total energy-momentum tensor
and integration is done only for the region where $u^r$ (radial
component of 4-velocity of fluid) is positive.
It is noted that the contribution of the rest mass energy is subtracted.
Factor 2 is coming from the symmetry of the system with respect to the
equatorial plane.
Blue curve represents the jet energy within the opening angle
    $\theta = 5^{\circ}$, while red curve represents the one within
    $\theta = 10^{\circ}$.  
The unit of vertical axis is $10^{48}$ erg. 
It is clearly seen that a more rapidly rotating black hole is driving
a stronger 
jet. The total energy of the jet for Model D ($a=0.95$) is as large as
$10^{50}$ erg. 
For comparison, 
the analytic curves of the BZ-flux formulation of 
Tchekhovskoy et al. (2010) for the case of monopole solution
B=$5 \times 10^{14}$G are shown by the red-dashed curve (until forth power of 
$\Omega_{\rm H} = a/2r_{\rm H}$ where $r_{\rm H}$ is the horizon of the 
black hole). 
The analytic curve of the BZ-flux formulation of Tanabe and Nagataki (2008) for 
the case of B=$5 \times 10^{14}$G is shown by the black-dashed curve (until forth power of $a$).
It is seen that the jet energy obtained in this study fits well with the 
previous analytical formulations.

\section{Discussion}\label{discussion}

1. Whole Picture.\\
In section \ref{result}, we could see that a stronger jet is driven by
a more rapidly rotating black hole. It is clear that the rotating
black hole is driving the jet. In fact, no jet is seen for the case
of Model A ($a=0$). On the other hand, the total energy of the jet for
Model D ($a=0.95$) is as large as $10^{50}$ erg. 
We showed some figures for Model D to understand
the dynamics more clearly. We have found that (i)
the stagnation region is seen around $r=15$ in the jet region, 
(ii) the ordered poloidal field lines are seen in the jet (that is
consistent with other previous study on the jet formation from a torus
(e.g. \cite{mckinney04})), (iii) the jet region is surrounded by
a 'Wall-like' structure that has a higher pressure than the jet region
and contains strong vertical magnetic fields, and (iv) the jet is
initiated by outgoing Poynting flux from the outer horizon.
effect (Blandford-Znajek effect).

2. Wall-Like Structure.\\
We found the 'Wall-like' structure around the jet.
It is considered that
the structure will depend on the initial angular momentum
and density profile of the progenitor star and initial configuration 
and its strength of magnetic fields. We should investigate this feature by 
performing some simulations systematically as a next step. 
However, this Wall-like structure seems not to work so much for the formation
and collimation of the jet, because the collimated jet has been launched already 
before the formation of the Wall-like strucuture. 

3. Jet Energy.\\
It has been shown that the jet energy can be as large as $10^{50}$ erg in
1.5 sec for Model D. Thus we expect that the energy of the jet can
be $\sim 10^{51}$ erg in 10 sec, and we can say that Blandford-Znajek
mechanism is very promising to drive a GRB jet. On the other hand,
for a slowly-rotating black hole case, the jet power looks too low, because
the luminosity of a GRB jet ($10^{51}$ erg s$^{-1}$) is hard to
explain. Thus it is concluded that a rapidly rotating black hole is
favored ('faster is better'). This result is 
consistent with some previous test calculations
\citep{tanabe08,tchekhovskoy10}.
Due to the limit of computational resources, we had to stop the
simulation at $t=160000$ in this study. We are going to present a simulation
that should last more than 10 sec in the near future.

4. Lorentz Factor.\\
It is found that the bulk Lorentz factor of the jet cannot be so high
in this study (of the order of unity), although 
the electro-magnetic field energy flux is much greater than rest-mass 
flux in the jet, which was
also seen in Nagataki (2009). We believe this jet has a good
property, because the jet speed will become relativistic if the
electro-magnetic field energy is converted to kinetic energy through
the propagation. Also, we can be optimistic because the Wall-like
structure was found in our simulations. It is pointed out that such a
Wall is necessary to drive a relativistic jet in some papers
\citep{komissarov09b, tchekhovskoy10}. Moreover, there have
been many papers on the propagation of relativistic jet from a massive
star~\citep{aloy00,aloy02,zhang03,zhang04,mizuta06,mizuta09,morsony07,wang08,morsony10,mizuta10,nagakura10}
where the jet is driven by injecting energy from the inner
boundary. Some of their results suggest that the bulk Lorentz factor
of the jet will be increasing with time (e.g. \cite{mizuta10}) because
the material of the progenitor star is blown off and the jet region
becomes 'cleaner' with time. Thus we expect that a
relativistic jet will be seen if we perform a simulation for a longer
time scale (of the order of $10-100$ sec. see also Suwa and Ioka 2010).

5. Blandford-Payne Effect.\\
In this study, we showed that Blandford-Znajek effect is working. 
On the other hand, it was found that no jet was driven with a help
of rotation of the accretion disk (Blandford-Payne effect) in 
Model A ($a=0$). Of course there is a possibility that the disk will
drive a jet later. Also, the interaction between the
rotating black hole and the accretion disk may be important. That is, 
if the rotating black hole is connected with the accretion disk by
magnetic fields and
rotation power is conveyed to the disk, Blandford-Payne effect will 
become more effective. Further study will be necessary to conclude the
necessity of Blandford-Payne effect to drive a GRB jet.

6. Microphysics.\\
In this study, microphysics such as neutrino physics and nuclear
reaction is not included. 
The disk structure is deformed by neutrino cooling process
\citep{popham99,nagataki02,nagataki03a,nagataki07,lopez10,linder10,taylor10,sekiguchi10}.
Nucleosynthesis study provides us with an opportunity to compare the
simuations with observations of supernova ejecta and supernova
remnants \citep{nagataki97,nagataki98a,nagataki98b,nagataki00,maeda02,nagataki03b,takiwaki04,nagataki06,maeda07,tominaga09,ramirez-ruiz10,milosavljevic10}. 
We are planning to include these microphysics as a next step.

7. 3D Simulation.\\
We showed our results of two-dimensional (2D) simulations in this study. 
It will be important to compare the 2D simulations with 3D ones. 
In 3D simulations, hydrodynamic instability in the azimuthal direction 
will be seen~\citep{nagakura08,nagakura09,mckinney09}.

\section{Conclusion}\label{conclusion}

We have performed 4 simulations of collapsars with different 
Kerr parameters $a=$0, 0.5, 0.9, 0.95. It is clearly shown that a more
rapidly rotating black hole is driving a more energetic jet. 
No jet is seen for the case of Model A ($a=0$), while the total
energy of the jet for Model D ($a=0.95$) is as large as $10^{50}$
erg. We expect that the energy of the jet can be $\sim 10^{51}$ erg in
10 sec for Model D. In order to explain the high luminosity of a GRB,
it is concluded that 'faster is better'. 

We have found in Model D that 
(i) the stagnation region is seen around $r=15$ in the jet region,  
(ii) the ordered poloidal field lines are seen in the jet,
(iii) the jet region is surrounded by the
'Wall-like' structure that has a higher pressure than the jet region
and contains strong vertical magnetic fields, and 
(iv) the jet is initiated by outgoing Poynting flux from the 
outer horizon (Blandford-Znajek effect). 

As for the bulk Lorentz factor, it is still not so high in this study 
(of the order of unity), although energy density of electro-magnetic field 
dominates the one of rest-mass in the jet. We expect
that a relativistic jet will be seen if we perform a
simulation for a longer time scale (of the order of $10-100$ sec).

\bigskip

This research was supported by Grant-in-Aid for Scientific Research on
Priority Areas No. 19047004 and Scientific Research on Innovative
Areas No. 21105509 by Ministry of Education, Culture, Sports,
Science and Technology (MEXT), Grant-in-Aid for Scientific Research (S)
No. 19104006 and Scientific Research (C) No. 21540404
by Japan Society for the Promotion of Science (JSPS),
and Grant-in-Aid for the Global COE Program
"The Next Generation of Physics, Spun from Universality and Emergence" 
from MEXT of Japan. The computation was carried
out on NEC SX-8 at Yukawa Institute for Theoretical Physics (YITP) in
Kyoto University and Cray XT4 at Center for Computational Astrophysics
(CfCA) in National Astronomical Observatory of Japan (NAOJ).


\clearpage

\clearpage

\end{document}